\documentclass{ifacconf}

\usepackage{graphicx}      
\usepackage{natbib}        
\usepackage{color}
\usepackage{amsmath}
\usepackage{amsfonts}
\usepackage{amssymb}

\graphicspath{{images/}}

\begin{document}
\begin{frontmatter}

\title{An Industrial Social Network for  Sharing Knowledge Among Operators\thanksref{footnoteinfo}} 

\thanks[footnoteinfo]{The research has been carried out within the ``Smart and adaptive interfaces for INCLUSIVE work environment" project (INCLUSIVE), funded by the European Union’s Horizon 2020 Research and Innovation Programme (H2020-FoF-04-2016) under grant agreement N.~723373.}

\author{Valeria Villani, Lorenzo Sabattini, Alessio Levratti, Cesare Fantuzzi}

\address{Department of Sciences and Methods for Engineering (DISMI), University of Modena and Reggio Emilia, Reggio Emilia, Italy\\(e-mail: name.surname@unimore.it).}

\begin{abstract}
	Due to the increasing complexity of modern automatic machines typically used in several industrial applications, the need for assistive technologies is becoming very relevant. Typical approaches consist in designing advanced and adaptive human-machine interfaces (HMIs) that can be effectively used by any operator and that provide guided procedures for the most common situations. However, when dealing with complex systems, infrequent and unforeseen situations may happen, whose solution require the experience owned by a limited number of skilled operators. To this end, in this paper we propose an industrial social network concept to allow an effective exchange of information among the operators and to facilitate the solution of unforeseen events, such as unscheduled maintenance activities or troubleshooting.
\end{abstract}


\end{frontmatter}

\section{Introduction}

In recent years market demands and advances in production technologies have been leading to a dramatic increase in the complexity of modern automatic machines. Indeed, while fast production rates with high quality cannot be missed, it is required that modern production systems implement other advanced functions, such as fault diagnosis and fast recovery, fine-tuning of process parameters, and fast reconfiguration of the machine parameters to adapt to production changes.
In spite of such requirements on flexibility and performance, such systems, and the plants where they operate, are in practice far from being automatic and autonomous, but still require the constant presence of human operators to control and supervise the basic and the advanced functions.

Human operators interact with such machines by means of human-machine interfaces (HMIs), which are the main (if not only) way an operator can achieve proper situation awareness about the current status of the machine and undesired interactions in production.
As a consequence of the above mentioned scenario, the increased complexity of production systems largely reflects in highly complex HMIs. In other words, the complexity of the machine is somehow passed to the human operators, who are requested to be more and more skilled to be able to operate such systems. On the one side, this clearly implies that an increasing, and almost unsustainable, burden is put on the operators, who are requested to supervise and interact with very complex systems, typically under challenging situational conditions, such as noisy environment, tight time constraints, fear of job loss, and/or psychological pressure due to the presence of supervisors. On the other side, novice or unskilled operators are almost unable to interact with such systems, unless they are provided proper training. This applies also to elderly (and middle age) workers who feel uncomfortable in the interaction with a complex computerized system, despite having a great experience about the underlying production processes.

To overcome these issues, in addition to a change in the design paradigm of HMIs that might lead to usable and anthropocentric interaction systems \citep{Nielsen_usability_engineering, Nachreiner_2006}, proper tools should be conceived to support operators during the interaction, specifically to assist them to overcome infrequent or unforeseen situations, such as unscheduled maintenance activities or troubleshooting. Indeed, these are the kinds of activities that are less likely for operators to come familiar with, occurring the former seldom, thus being not memorable, and the latter under the effect of anxiety and strict time constraints.

\begin{figure}
	\centering
	\includegraphics[width=\columnwidth]{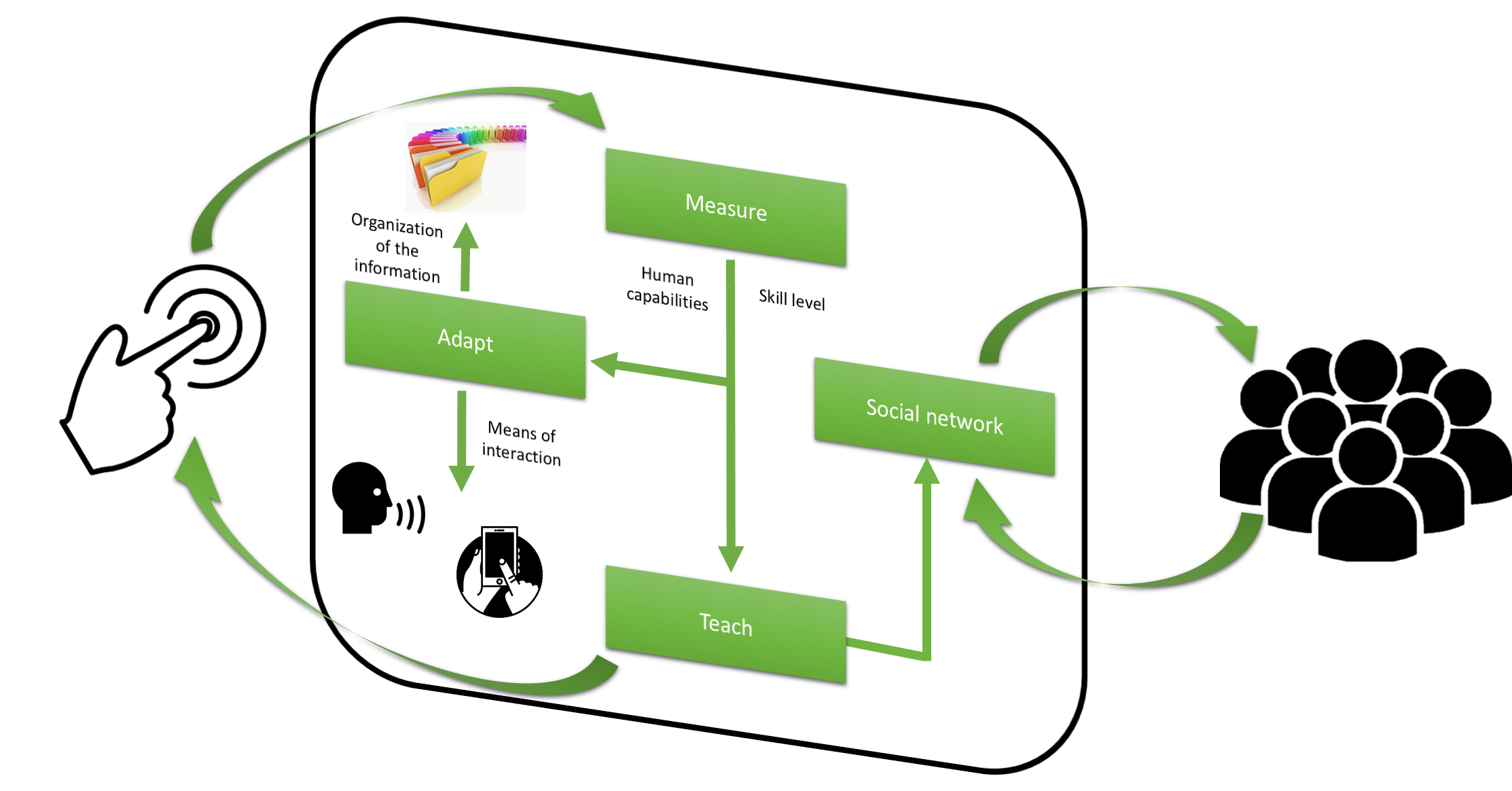}
	\caption{\label{fig:overview_inclusive}Main components of the INCLUSIVE system.}
\end{figure}

In this paper, we propose the concept of an industrial social network as a peer-to-peer tool for sharing knowledge among shopfloor operators and complement other existing support tools, if any. Specifically, the industrial social network is conceived in the framework of the European project INCLUSIVE, which aims at developing smart and adaptive interfaces for inclusive work environment in industries \citep{Villani_2017_ETFA}. To achieve this, the INCLUSIVE system is organized in three main pillars, which are shown in Fig.~\ref{fig:overview_inclusive}. The first pillar relates to the measurement of human capabilities: the system will measure the human capability of understanding the logical organization of information and the cognitive burden the operator can sustain. The second pillar consists in the adaptation of interfaces to the measured human capabilities. Finally, the third pillar is about teaching and training for unskilled users: the system will be able to teach the correct way to interact with the machine to the unskilled users, exploiting also simulation in virtual (off-line training) and augmented (on-line training) environment.

The industrial social network is meant as a support tool and a further complement to training provided with the aid of virtual and augmented reality. Specifically, details about how to use the system, what its current status is and how to address any occurring faults will be provided with virtual and augmented tools. However, these will likely be unable to cover all possible needs of guidance for every operator, given the unpredictability of possible operators' action, error situation and, ultimately, machine faults. In these scenarios, the use of a social network devoted to the working environment would be advantageous to solve the issue by quickly reaching those colleague(s) most competent on the topic. Ultimately, the support provided by such a proposed tool is a further step towards the inclusion of any workers and the improvement of the accessibility of complex production machines, which in turn allow for an increase in the overall system efficiency and productiveness.

The paper is organized as follows. First, the related work on support and training tools is recalled in Sec.~\ref{sec:related_work}. The proposed concept for an industrial social network is presented in Sec.~\ref{sec:concept} and Sec.~\ref{sec:sw_arch} describes the preliminary software architecture. Some final remarks are discussed in Sec.~\ref{sec:conclusion}.

\section{Related work}\label{sec:related_work}
While in industrial practice manuals are still largely the most widespread tool supporting operators in the management of critical scenarios, such as errors, alarms and faults, it is nowadays well known that they suffer from a huge number of problems that make their use ineffective and inefficient in shopfloors \citep{Crowder_2003, Setchi_2003}. As a first improvement, the use of hypermedia information systems, usually named hypermedia maintenance manuals, industrial hypermedia applications or interactive electronic technical manuals has been proposed \citep{Malcolm_1991, Setchi_2003, Crowder_2003, Greenhough_2007, Fakun_2002}. These allow efficient information retrieval, thus facilitating the task-oriented structuring of information sources in technical manuals and the extraction of specific information. 
Their primary goals are supporting the activities of the maintenance personnel in performing preventive and corrective maintenance \citep{Setchi_2003}, troubleshooting \citep{Villani_2016_HMS_MyAID} and allowing the integration of resources across an enterprise through information and document maintenance \citep{Crowder_2003}. 

Furthermore, systems specifically devoted to training operators have been introduced recently. These mainly rely on the use of virtual reality technologies, which allow operators to be trained without physical equipment or trainer involvement, thus reducing costs and machine downtime. Additionally, virtual training allows to train without being exposed to the inherent risks of the task \citep{Liang_2012}, or to train infrequent incidents, for instance in construction projects \citep{Petridis_2009}.
For example, in \citep{Ordaz_2015} a virtual simulation and training system is presented that aims at training operators to perform manual tasks for automotive manufacturing. In \citep{Loch_2017} an adaptive virtual training system for machine operation that is addressing the ageing workforce is considered. Focus is given to the didactic approach and the interaction modalities for the specific class of target users, namely ageing employees.
However, although several studies have confirmed the positive impact of virtual training on procedural learning in manual industrial tasks \citep{Adams_2001, Lin_2002, Gorecky_2017}, virtual training has not become a standard in daily industrial practice yet. This is mainly due to limited users' acceptance and the authoring efforts for collecting the relevant data, defining the training scenarios and plans \citep{Gorecky_2017}.
These drawbacks are partially solved by on-line support tools relying on augmented reality, which assist and guide workers during the execution of manual tasks. These systems augment the reality that the user is viewing since graphical instructions are displayed that indicate the working steps. As a consequence, the most promising uses of augmented reality in industrial applications are related to design, assembly and maintenance since they allow to display synoptic information onboard the manufacturing system and in the field of vision, such as performance values, catalogue spare part codes and work instructions \citep{Michalos_2016, vanKrevelen_2010}. Tasks such as assembly design, where the optimal assembly operation sequence that minimizes completion time and effort must be found \citep{Ong_2007}, and guidance \citep{Yuan_2005} can be enhanced. The added value of the use of augmented reality in these operations is that proper real-time training is given to operators, providing them with cues and guidance. The same applies in the case of maintenance tasks, where real-time troubleshooting and spare parts purchase actions with all relevant information and functions are enabled to maintenance personnel \citep{Nee_2012}.
However, training tools based on augmented reality, as well as virtual training, require that the training scenarios are defined a priori and all possible users' needs should be identified in the design phase, which is quite impossible to achieve. Thus, despite providing useful and intuitive support for the decoded scenarios, training by means of virtual and augmented reality intrinsically cannot provide complete assistance and guidance to human workers.

Moving along these lines, the proposed concept of industrial social network is intended to complement and complete troubleshooting documentation and other interactive teaching tools, by addressing users' needs and errors and working scenarios that cannot be predicted preliminarily. This is achieved by relying on today's people acquaintance with using social network, searching for solutions to everyday problems on the internet and consulting online discussion forums. This is expected to favour the user's acceptance of such a working support tool.

\section{Proposed concept}\label{sec:concept}


The industrial social network is meant as an application running on handheld devices, such as standard smartphones or tablets (e.g. those running iOS or Android), supporting workers when they cannot solve the problems they are facing, despite the presence of other teaching and training support tools. 
In particular, the industrial social network can be seen as a virtual board where users post the current machine problem, enriched with possible media (e.g., pictures, videos, text messages from the machine).
The request is either automatically or manually sent to the local workers community or the service level who has the app installed on her/his mobile device. Filters on the addressees are set so that only those workers who can actually contribute to the discussion, given their expertise and knowledge, are addressed. Thus, great importance should be given by management and human resources staff to defining clusters of competences and expertises across the company and mapping workers' skills to them. Further, for a precise identification of the skills required to solve a fault situation, metadata associated to alarms and troubleshooting procedures could be considered. Thus, the user can decide to send the message to all the other workers, or a single person, or a subset of workers, selected by their experience level.
Additionally, by exploiting any possibly available landmarks, the industrial social network app can also locate the position of each user in the plant. This information can be used to select, among the addressees of a message, those who are located in the proximity of the worker seeking for assistance, provided that the required skills are available.

Once a message has been posted on the social network, the reached operators can declare their availability to provide support and a direct link can be established among the sender and the willing colleagues, in the form of a private chat or a (video-)call. If no availability is provided, the thread will be still available to all the reached workers, but they will not be notified online about the ongoing discussion.


\subsection{Use case}

\begin{figure}
	\centering
	\includegraphics[width=\columnwidth]{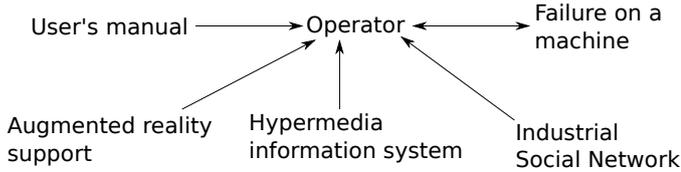}
	\caption{\label{fig:schema_use_case}Scenario of a possible use case.}
\end{figure}

\begin{figure*}
	\centering
	\includegraphics[width=\textwidth]{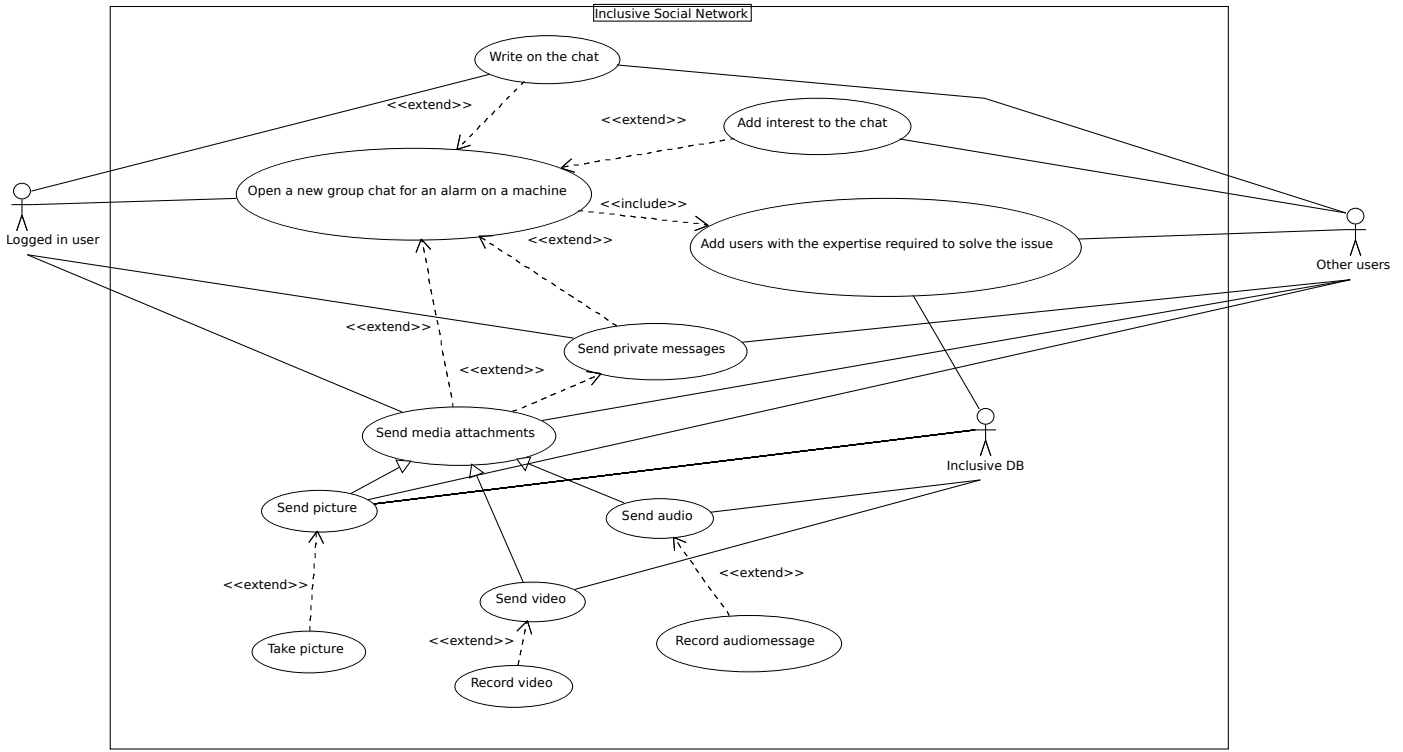}
	\caption{\label{fig:usecasediag}Use case diagram.}
\end{figure*}

As a possible use case of the proposed concept of an industrial social network, let us consider the one depicted in the use case diagram depicted in Fig.~\ref{fig:usecasediag}. Specifically, we consider an operator, denoted in the following by ${OP}_A$, working at an automatic machine. As depicted in Fig.~\ref{fig:schema_use_case}, we consider that, in addition to the social network, the operator is provided with some support and troubleshooting tools, such as the machine manual, a hypermedia information system and an on-line support tool based on augmented reality.

In this scenario, let us denote the possible levels of experience in the company as the set $E=\{{exp}_1, {exp}_2, {exp}_3, \dots\}$, where $exp_{i+1}$ denotes an higher level of experience than $exp_i$ and, hence, enables access to a greater set of duties to operate the machine. Finally, without any loss of generality, we assume that ${OP}_A$ has experience $exp_{2}$.

At a given time instant, ${OP}_A$ is prompted with an alarm occurring at the machine (s)he is working with. The HMI shows a short description of the alarm, which is insufficient for ${OP}_A$ to solve the alarm. Moreover, no support is provided in this regard by the other available support tools, e.g., troubleshooting section of the hypermedia information system of the machine and training based on augmented reality. Moreover, the description of the alarm under consideration states that the alarm can be solved by workers whose level of experience is $exp_{5}$ or greater.
Thus, the industrial social network is the last resort for ${OP}_A$ to restore the machine.
${OP}_A$, then, writes a message on the virtual board of the industrial social network, attaching a picture of the HMI describing the alarm. Using a filtering function, (s)he selects all the workers who have level of experience $\geq exp_{5}$, are currently in the plant and are close to the machine where (s)he is working.
Clearly, all the workers satisfying the filtered features are notified of the new thread. Among them, operators ${OP}_B$ and ${OP}_C$ show their availability to help ${OP}_A$. A dedicated chat is set up among ${OP}_A$ and all the workers who are available to support her/him. The members of this chat are continuously updated as soon as a new worker declares her/his availability.
Ultimately, by using this chat, ${OP}_A$ gets the information and/or assistance needed to solve the fault.

\subsection{Functional and implementation requirements}
From the presented use case, a set of functional and technical specifications and requirements can be defined, irrespective of the specific requirements of the application.
Functional specifications mainly refer to the need to find a consistent representation of needed and available skills, so that a good match can be established between operators seeking for help, and those who could provide it.
Specifically, on the one side, this refers to the importance of achieving a convenient and exhaustive synthetic representation of skills and expertises available among all workers.
On the other side, it is needed to collect all the information about competences and skills required for alarms, faults and procedures that is already available in the shopfloor (e.g., in manuals). This becomes fundamental for an effective automatic filtering of operators that might be interested in a new thread.

From the technical point of view, the industrial social network should have access to the device camera, speakers and microphone. Thus, it should allow to post short text messages and attach pictures, videos and voice messages. Moreover, the app should allow to localize users in the plant, at least if a localization infrastructure is available. Finally, as another main technical feature, an important part of the social network is the database of workers, which must collect in a complete, but still structured and organized form, all the workers and their expertises and competences, as discussed above. Moreover, such database should retain information about which operators are currently present in the plant.

%

\section{Industrial social network architecture}\label{sec:sw_arch}
\subsection{System overview}
\begin{figure}
	\centering
	\includegraphics[width=\columnwidth]{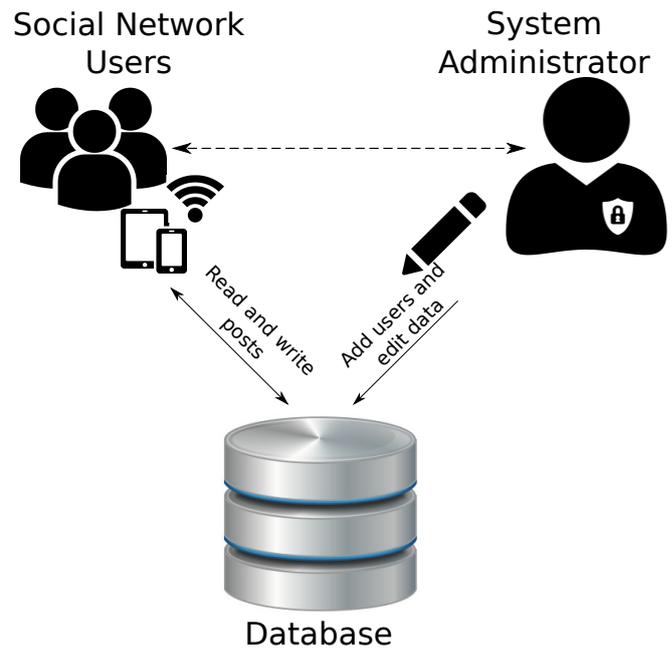}
	\caption{Architecture of the industrial social network: data storage and collection on a database.}
\end{figure}

The industrial social network is designed as a web based application: devices running the application need to be connected to internet either by a Wi-Fi or mobile network. In the case the company exploiting the application's services wishes to keep data reserved, devices can be connected to the company network. Then, in order to share and collect data regarding the issues, the app needs to exploit database services. The database can be stored using cloud services or locally, in the company's servers. 

For the correct usage of the application, users must share some information about their job and about their working status. In particular, users should provide data regarding the following fields: 
\begin{itemize}
	\item name and surname of the user,
	\item email (needed for the user's authentication during login as well as a personal and unique password),
	\item experience level of the user,
	\item expertises (i.e. programming capabilities, skills in using tools, etc\dots) acquired while working or by following courses,
	\item duties usually assigned to the user, intended as competences in using particular machines.
\end{itemize}
These data are recorded and modified only by the system administrators of the application and should not be modified by the users. Furthermore, for privacy and data protection reasons, only the system administrators can have access to the database storing the data. 

Each post published on the social network is also stored into the database. As complementary information, the post must be attached also with data about the error code generated by the machine, the duty associated with the machine and a list of the expertises required to solve the issue. The post should also include a clear and synthetic title and a description of the issue, in order for the other users to understand the problem. The post can also be accompanied with some multimedia material, such as videos, pictures or vocal messages, to further clarify the issue on the machine and provide any useful information to other users. 

The application, that runs on user's tablet or smartphones, retrieves from the database data associated with the authenticated user during the login activity and autonomously filters the multiple posts published on the social network according to data provided while creating new posts.

In order to avoid bothering busy users, once a new post has been published, users can decide whether to accept or decline to contribute in the discussion on solving the issue. This functionality is also useful for users who do not consider themselves capable on positively contributing to the discussion.

\subsection{Software architecture}

\begin{figure*}[htbp]
\centering
	\includegraphics[width=\textwidth]{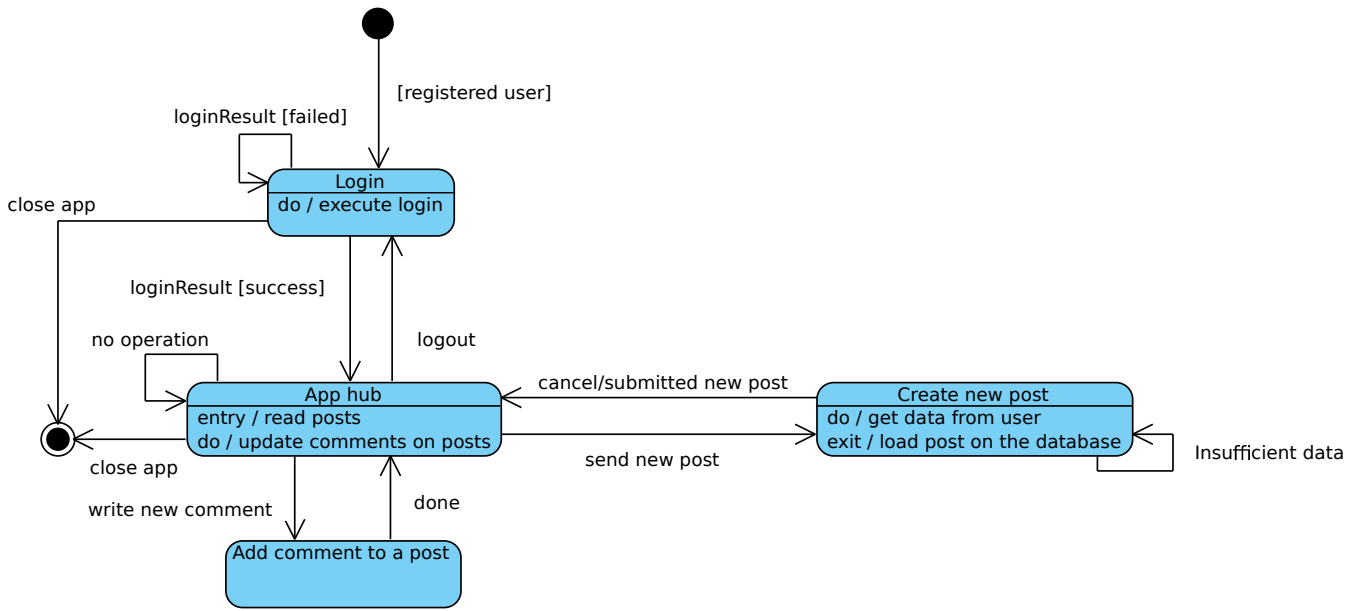}
	\caption{UML 2.0 compliant State Machine diagram representing the Inclusive Social Network application}\label{fig: stateChart}
\end{figure*}
Figure \ref{fig: stateChart} reports a simplified state chart of the implementation of the industrial social network application. 

User's personal data, such as expertises, duties and experience level, are managed by the system administrator, thus, registration is not possible for simple users: only the administrator is allowed to create a new account. The application, then, starts with the login (``\textit{Login}'' state) for existing users. Each user must login by providing their email and password. Once login has been successfully executed, the app leads the user to the application's hub (``\textit{App hub}'' state). While in this state, the application queries the database for new posts and filters them according to user's data, provided during the registration, and posts data recorded during the creation of a post (``\textit{Create new post}'' state). In this state, users can create a new post on the social network by specifying the following mandatory information:
\begin{itemize}
	\item title of the post,
	\item error code shown by the machine the user is working at,
	\item duty or machine's name that generated the alarm / error / issue,
	\item required expertises for solving the issue,
	\item receiver (optional, in order to create private chats),
	\item body of the message, where users should describe the issue, the operations executed to generate the issue and, possibly, multimedia files (including videos, pictures or vocal messages) in order to best describe the issue. 
\end{itemize}

It is possible for the user to publish the post, which is then loaded on the database, if and only if the user provides all the aforementioned data (except for the receiver of the post that is optional). After the publication of the post, and its recording on the server, the application leads back the user to the hub, where the newly published post appears. While in the ``\textit{App hub}'' state, the application shows the filtered posts to the users. From this state, users can select a visible post and add comments to it.

\subsubsection{Example}
The following represents a typical example of use of the industrial social network. Operator ${OP}_A$, who is a beginner in PLC programming and has a low experience level, receives an alarm on a \textit{cartoning machine} (that is as well registered on the database as a machine of the company) inherent to a PLC programming (a company's job inherent expertise) error that generates an unhandled exception. Operator A, then, publishes a post by filling the form for the creation of a new post with the name of the machine (s)he is working with (in this case the Cartoning machine) and the \textit{PLC programming} expertise that is required to solve the programming error. Operator ${OP}_B$, whose duties include working with the cartoning machine, and operator ${OP}_C$, who is a skilled PLC programmer, receive a notification for a new post on the social network. Then, by selecting the new post, written by ${OP}_A$, the app leads them to the state ``\textit{Add comment to a post}'', where they can give help to solve the problem. 

\section{Conclusion}\label{sec:conclusion}

In this paper we proposed a novel concept of an industrial social network, to be used for knowledge sharing among operators in industrial environments. The industrial social network is intended as a complement to advanced and adaptive HMI systems, to provide assistance to the operators when unforeseen and infrequent situations take place. In particular, colleagues with sufficient knowledge will be notified of the situation, and will be allowed to provide assistance to the operator in need, sharing their knowledge. 

The general architecture of the industrial social network system was proposed, together with a preliminary software architecture implementation.

Current work aims at finalizing the implementation of the industrial social network application, and its evaluation in real operative conditions, exploiting the industrial use cases provided in the framework of the INCLUSIVE European project.


\bibliography{INCLUSIVE_INCOM2018}             
                                                   







\end{document}